\documentclass[conference]{IEEEtran}
\IEEEoverridecommandlockouts
\usepackage{cite}
\usepackage{amsmath,amssymb,amsfonts}
\usepackage{amsthm}
\usepackage{algorithmic}
\usepackage{graphicx}
\usepackage{textcomp}
\usepackage{xcolor}
\graphicspath{ {/home/user/Images/} }
\DeclareMathOperator{\Prob}{Prob}
\newcommand{\coloneq}{\mathbin{\raisebox{.32pt}{:}{=}}}

\theoremstyle{definition}
\newtheorem{definition}{Definition}[section]

\theoremstyle{remark}

\makeatletter
\newcommand*{\rom}[1]{\expandafter\@slowromancap\romannumeral #1@}
\makeatother
\def\BibTeX{{\rm B\kern-.05em{\sc i\kern-.025em b}\kern-.08em
    T\kern-.1667em\lower.7ex\hbox{E}\kern-.125emX}}
\begin{document}

\title{From Bits to Insights: Exploring Network Traffic, Traffic Matrices, and Heavy-Tailed Data}

\author{\IEEEauthorblockN{Christopher Howard, Hayden Jananthan, Jeremy Kepner}
\IEEEauthorblockA{\textit{Massachusetts Institute of Technology}}
}
\IEEEoverridecommandlockouts
\IEEEpubid{\makebox[\columnwidth]{979-8-3503-0965-2/23//\$31.00~\copyright2023 IEEE \hfill}
\hspace{\columnsep}\makebox[\columnwidth]{ }}
\maketitle
\IEEEpubidadjcol

\begin{abstract}
    With the Internet a central component of modern society, entire industries and fields have developed both in support and against cybersecurity. For cyber operators to best understand their networks, they must conduct detailed traffic analyses. A growing recognition is the ubiquity of heavy-tailed characteristics in network traffic. However, a thorough analysis of cybersecurity programs suggests little statistics educational background, worsened by the observation that college-level statistics courses largely lack heavy-tailed content, meaning cyber operators are both ill-equipped to appropriately analyze their network traffic and unable to easily access resources that could help. In response, we developed an accessible Jupyter Notebook module that guides individuals---regardless of statistical background---through traffic matrix creation, heavy-tailed data identification, data visualization, and distribution fitting. Such content empowers cyber operators, improving analyses and design.
\end{abstract}

\begin{IEEEkeywords}
    heavy-tailed statistics, traffic matrix, cybersecurity
\end{IEEEkeywords}

\let\thefootnote\relax\footnotetext{Research was sponsored by the United States Air Force Research Laboratory and the Department of the Air Force Artificial Intelligence Accelerator and was accomplished under Cooperative Agreement Number FA8750-19-2-1000. The views and conclusions
contained in this document are those of the authors and should not be interpreted as representing the official policies, either expressed or implied, of the Department of the Air Force or the U.S. Government. The U.S. Government is authorized to reproduce and distribute reprints for Government purposes notwithstanding any copyright notation herein.}

\section{Introduction}

``What is happening on my network?'' is the most critical question that any cyber operator could ask and, ideally, should be able to answer. After analyzing more than 85 programs across NSA National Centers of Academic Excellence in Cybersecurity (NCAE-C) with the `Cyber Research' designation as well as courses at some of the nation’s top actuarial science programs, we confirmed the pervasive absence of heavy-tailed curricula. Thus, a complete and satisfying answer to our motivating question remains out of reach to most cyber operators as fully understanding network behavior is impossible without understanding the heavy-tailed characteristics of the network traffic.

A key feature of heavy-tailed distributions is the `heaviness' of their tails, i.e., that their tails decay especially slowly, the watermark of which is that of exponential distributions. This slow decay implies that events described by a heavy-tailed distribution are more likely to take on extreme values---deviations far from the mean---when compared to their light-tailed counterparts \cite{wierman2022fundamentals}. This makes them highly suited to describing the statistical patterns of network traffic which exhibit bursts of high traffic between source/destination pairs; while such extreme bursts are rare, they \emph{do} occur with higher regularity than a light-tailed distribution like a Gaussian model would suggest. Consequently, heavy-tailed distributions often best model the behavior of network traffic \cite{kepner2019hypersparse}.

To teach network operators more about heavy-tailed distributions, we have designed an interactive Jupyter Notebook lab using D4M \cite{kepner2012dynamic} to analyze network traffic data from the WIDE group \cite{cho2000traffic}, with the ultimate goal for users to better understand network traffic, traffic matrices, and heavy-tailed data.

The format for the paper is as follows: \S\ref{background section} briefly describes the history and role of heavy-tailed statistics, including a formal definition of what it means for a distribution to be heavy-tailed. \S\ref{analysis section} focuses on the analysis of cybersecurity programs at NCAE-C `Cyber Research' institutions and overall heavy-tailed statistics education. \S\ref{lab section} describes the various components of the constructed Traffic Matrix Lab and the design and teaching philosophy used in its creation.

\section{Background}
\label{background section}

As mentioned above, a key feature of heavy-tailed distributions is the slow decay of their tails, particularly when compared to exponential distributions. More formally, we follow the approach used in \cite{wierman2022fundamentals} by defining heavy-tailed distributions in terms of their complementary cumulative distribution functions.

\begin{definition}[complementary cumulative distribution function (CCDF)]
    The \emph{complementary cumulative distribution function} (CCDF) $\overline{F}_X$ of a continuous random variable $X$ supported on an interval $I \subseteq \mathbb{R}$ is defined as: 
    \begin{equation*}
        \overline{F}_X(x) \coloneq \Prob(x \leq X) \qquad (x \in I).
    \end{equation*}
\end{definition}

\begin{definition}[heavy-tailed distribution]
    A random variable $X$ supported on an interval $(a, \infty) \subseteq \mathbb{R}$ is \emph{heavy-tailed} if its tail decays more slowly than that of any exponential distribution (which have CCDFs of the form $e^{-\mu x}$), i.e., if for all $\mu > 0$,
    \begin{equation*}
        \limsup_{x \to \infty}{\frac{\overline{F}_X(x)}{e^{-\mu x}}} = \infty,
    \end{equation*}
    or, equivalently, for every $x_0 > a$ and every $M > 0$ there exists $x > x_0$ such that $\overline{F}_X(x)/e^{-\mu x} > M$. 

    A random variable $X$ is \emph{light-tailed} if it is not heavy-tailed. 

    A distribution is \emph{heavy-tailed} or \emph{light-tailed} if any (equivalently, all) random variables having that distribution are heavy-tailed or light-tailed, respectively.
\end{definition}

Although statistics education tends to focus almost exclusively on light-tailed distributions and statistical techniques based on those distributions, heavy-tailed statistics does maintain a strong history. In the 1890s, Italian economist Vilfredo Pareto observed that approximately 80\% of Italy's land was owned by 20\% of the population \cite{pareto1971translation,chakrabarti2004rmat}; this same phenomena was recorded repeatedly elsewhere before eventually leading to the creation of the Pareto Distribution. The Pareto Distribution, one of the most famous heavy-tailed distributions, both captured the wealth disparity seen in many nations and inspired the 80/20 rule (80\% of all outputs comes from 20\% of all inputs) \cite{pareto1971translation}. Other real-world examples of heavy-tailed distributions at play include social network activity \cite{chakrabarti2004rmat,barabasi1999emergence}, epidemic outbreaks \cite{maji2021identifying}, and, most topically, internet traffic \cite{kepner2019hypersparse}. However; despite the many real-world applications of heavy-tailed statistics, to most individuals the emergence of a heavy-tail is viewed as an enigma rather than a predictable outcome. 

While the relatively slow tail decay rate certainly makes heavy-tailed distributions special, that is not the only unique feature that these distributions possess. Another unique feature of these distributions is the absence of well-defined moment-generating functions \cite{wierman2022fundamentals}; when working with light-tailed distributions all moments are finite, but heavy-tailed distributions instead often have infinite moments (such as infinite variance). This means that higher-order moments like the variance, skewness, and kurtosis may not be well-defined due to the abundance of extreme values. As such moments are often used in traditional statistical data analysis methods (typically geared towards working with light-tailed data), the possible nonexistence means that these techniques become unusable \cite{devlin2021hybrid}; consequently, characterizing a heavy-tailed distribution often proves to be more difficult than in the light-tailed case. Similarly, the absence of these higher order moments leads to the failure of the Central Limit Theorem, as one of its hypotheses is finite variance. 

Additionally, the high—or even infinite—variance of heavy-tailed data frequently leads to data points being separated by several orders of magnitude. As a result, it is typical to graphically represent heavy-tailed data on a log-log scale as well as logarithmically bin the data \cite{devlin2021hybrid}. Logarithmic binning is especially useful in highlighting tail behavior as one may easily take note on how the probability density decreases throughout the tail, as well as identify the extreme values influencing the distribution's properties. Another benefit of using the log-log scale is that it can aid in recognizing power law models (which are linear on a log-log scale).

\section{Analysis of Cybersecurity Programs}
\label{analysis section}

Empirical observations suggested that heavy-tailed education material present in cybersecurity and statistics courses was virtually nonexistent. To test this hypothesis, over close to 80 cybersecurity and statistics courses across 11 institutions and over a dozen online courses were examined for inclusion of heavy-tailed statistics material. The institutions looked at were drawn from NCAE-C `Cyber Research`-designated colleges and those with top actuarial programs while the online courses were sampled from the Class Central online courses aggregator. The motivation behind focusing on NCAE-C `Cyber Research'-designated institutions was the thought process that those institutions would require the most technically rigorous courses for training cyber operations. The emphasis on actuarial programs was borne from the fact that heavy-tailed distributions have application in risk analysis.

\begin{figure}[h!tbp]
    \centering
    \includegraphics[width=1.00\linewidth]{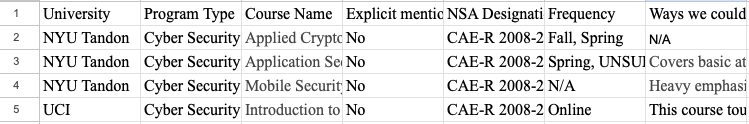}
    \caption{A depiction of a small subset of the data captured when analyzing the cybersecurity programs. The columns (left to right) read: `University', `Program Type', `Course Name', `Explicit Mention of Heavy-Tails', `NSA Designation', `Frequency', and `Ways We Could Implement A Discussion of Heavy-Tailed Distributions'.}
    \label{fig:enter-label}
\end{figure}

To conduct the analysis, statistics courses and cybersecurity courses were sampled from NYU Tandon, University of California Irvine, MIT, Dakota State University, Temple University, The University of Iowa, University of California Santa Barbara, the University of Connecticutt, University of Illinois Urbana-Champaign, and the University of Wisconsin-Madison. If the school had a cybersecurity program, all classes required for the degree were analyzed. If not, the statistics programs offered by the school were thoroughly investigated. For each school, information on the program a given course is taught within, course content, frequency that the course is offered at, NSA designation (if applicable), whether the course explicitly mentions heavy-tailed distributions, and where a discussion on heavy-tails could be implemented into the course was documented. Of the ninety-two courses that were analyzed, only three covered content pertaining to heavy-tailed distributions. Then, of those courses that did cover heavy-tailed distributions, one course---MIT course 1.151 ``Probability and Statistics in Engineering''---was not a regularly taught course and had not been offered since Fall 2018.
% and was not taught each year when it was offered by the department. 
Thus, approximately 2\% of the sample contained content related to heavy-tailed distributions and the initial hypothesis that heavy-tailed distributions were not taught not only in cybersecurity, but across most technical disciplines was confirmed.

\section{Traffic Matrix Lab Construction}
\label{lab section}

After verifying our initial hypothesis that there is an inadequate amount of educational material covering heavy-tailed distributions, the logical next steps were to design content to address those shortcomings. Drawing inspiration from \cite{jananthan2022youtube}, which leverages the Jupyter notebook environment \cite{kluyver2016jupyter} for visualization of network traffic statistics and heavy-tailed modeling, it was chosen that an interactive Jupyter notebook would be the vehicle of delivery, focused primarily on the traffic matrix approach to getting a holistic view of a network's traffic behaviors. Cyber operator students will use the Dynamic Distributed Dimensional Data Model (D4M) package \cite{kepner2012dynamic} to extract relevant traffic data from a subset of the MAWI dataset \cite{kepner2022new}, compute various network quantities, and exhibit the heavy-tailed nature of those quantities. 

\subsection{Design}

A major component of the educational philosophy employed while building the Traffic Matrix Lab is the Backward Design philosophy of Grant Wiggins and Jay McTighe in \textit{Understanding By Design} \cite{wiggins2005understanding}. `Backward Design' is a framework for designing content where curriculum planners first determine desired outcomes, then create assessments which determine of those desired outcomes have been achieved, and only then build lesson content with those assessments in mind. In designing the outline for the Traffic Matrix Lab it was decided that the Learning Objectives for the Lab were for cybersecurity professionals to understand heavy-tailed data, learn to identify heavy-tailed data, and learn to visualize that data using appropriate, modern statistical tools. 

\begin{figure}[h!tbp]
    \centering
    \includegraphics[width=\linewidth]{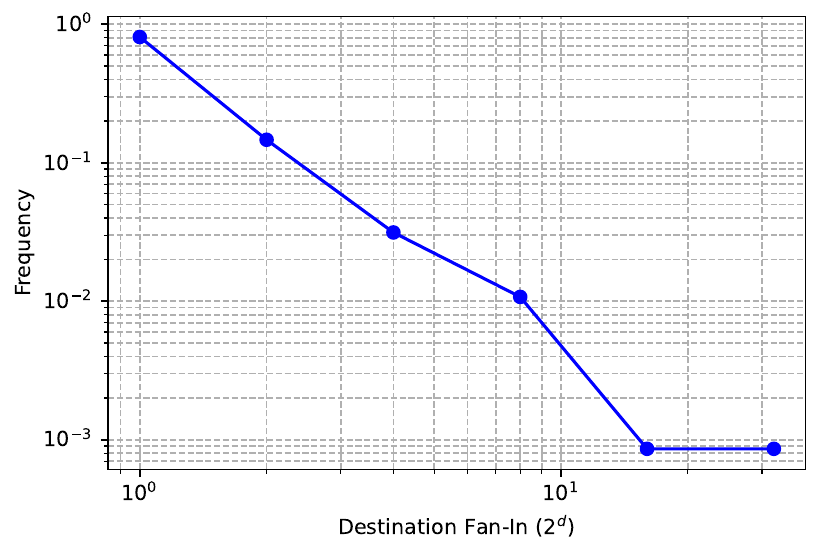}
    \caption{Frequencies of destination fan-in quantities among all destinations present in the MAWI dataset. Logarithmic binning is used, binning the actual destination fan-in $n$ to the nearest power of $2$, $2^d \leq n < 2^{d+1}$.}
    \label{fig:heavy-tailed-distribution}
\end{figure}

With the Learning Objectives outlined, the next concern became determining how to assess students' learning given the lab's asynchronous, online nature. Anticipating that the asynchronous engagement of students in open-ended lessons could become low over the course of the Lab, the Lab features several Knowledge Checks where students are asked relevant questions, checking their solution against provided sample solutions. As an explicit example, students were guided through plotting the frequency distribution of the (logarithmically binned) destination fan-in statistics across destinations, i.e., the number of sources that communicated with each destination. After the plot,  found in Figure~\ref{fig:heavy-tailed-distribution}, is constructed, students are asked various questions about the behaviors of the network which can be observed given the plot, such as: ``Are there any outliers that you see? How might these outliers impact the overall analysis of the network traffic?'' The given sample solution for this particular Knowledge Check is, ``The rightmost bin does not seem to follow our hypothesized power-law relationship. This may lead to a distorted interpretation of the scaling as power-law relationships rely on consistent trends across the data, and these outliers may disproportionately influence the line's slope.'' 

\begin{figure}[h!tbp]
    \centering
    \includegraphics[width=0.75\linewidth]{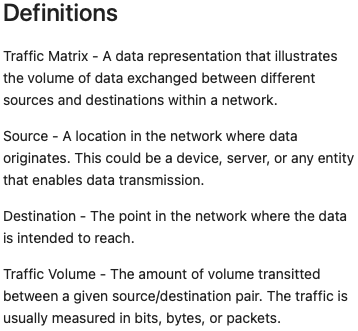}
    \caption{An abbreviated list of terminology and corresponding definitions within the `Definitions' section located at the introduction to the Traffic Matrix Lab.}
    \label{fig:definitions}
\end{figure}

Following the Backward Design approach, the final concern in designing the Lab was deciding what content \textit{must} be included for users to take away the key concepts. The MAWI dataset used for the lab features more than 84,000 rows of anonymized network traffic data held in a CSV file \cite{kepner2022new}. To appropriately analyze such a large dataset, it is important that users familiarize themselves with the Dynamic Distributed Dimensional Data Model (D4M) package \cite{kepner2012dynamic} as other data analysis packages struggle to carry out the same computations with comparable performance. Thus, it was necessary to include line-by-line explanations of what the code is doing so that users might generate similar code to analyze their own network traffic data. To assist those unfamiliar with traffic matrices, the lab also includes a `Definitions' section (see Figure~\ref{fig:definitions}) to ensure that users understand the terminology used throughout the Lab. Then, to emphasize the difference between heavy-tailed and light-tailed distributions, the Lab features a `Data Fitting' section where users can see the shortcomings of attempting to fit a light-tailed distribution to heavy-tailed data (see Figure~\ref{fig:light-tailed-fitting}). 

Section~1 of the Traffic Matrix Lab consists of a brief outline of course content followed by the aforementioned `Definitions' section. Section~2 of the Lab is centered on creating the relevant traffic matrices using D4M. There, users learn to extract different network quantities from a sparse traffic matrix $\mathbf{A}_t$ at time $t$. In particular, users calculate from the traffic matrix the `Destination Fan-In' (number of sources that communicate with each destination) and `Source Fan-Out' (number of destinations communicated with by each source) network statistics \cite{devlin2021hybrid}. In Section~3 users learn about methods and tools to visualize these network theoretic quantities and identify patterns in the data. Finally, Section~4 of the Lab focuses on fitting the distributions constructed in Section~3 to both light-tailed and heavy-tailed distributions, comparing their efficacy.

\subsection{Network Theoretic Quantities}

\begin{figure}[h!tbp]
    \centering
    \includegraphics[width=1.00\linewidth]{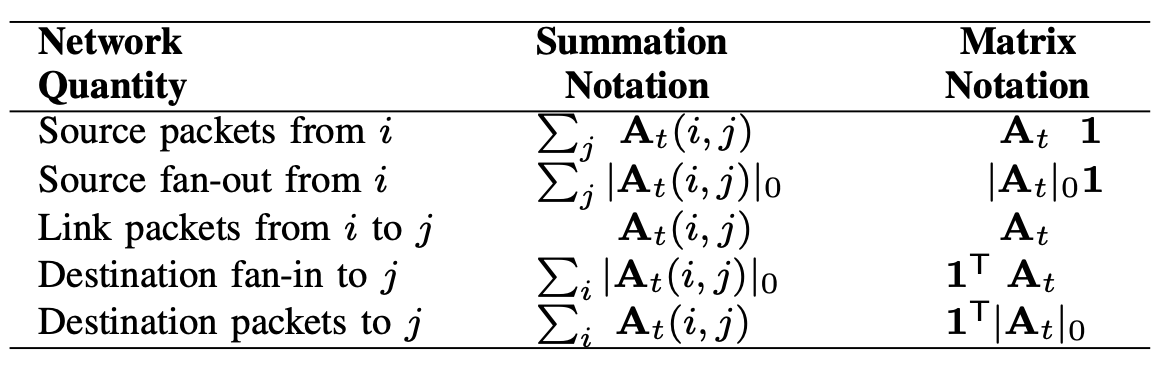}
    \caption{Short list of source- and destination-level network theoretic quantities along with mathematical expressions used to calculate them, adapted from \cite{kepner2019hypersparse}. Here $\mathbf{A}_t$ is the traffic matrix at time $t$, with rows and columns corresponding to sources and destinations, respectively. $\mathbf{1}$ represents an appropriately sized column vector consisting entirely of $1$s. $|\cdot|_0$ is the zero-norm, which when applied to a matrix replaces all nonzero entries with $1$ and leaves zero entries alone.}
    \label{fig:network_quantities}
\end{figure}

As previously mentioned, `source fan-out' refers to the number of distinct destinations to which a specific source sends data packets while `destination fan-in' refers to the number of distinct sources from which a specific destination node receives packets (Figure~\ref{fig:heavy-tailed-distribution}). These two network theoretic quantities have been extensively researched and shown to consistently follow a power-law relationship \cite{devlin2021hybrid}, which can be directly observed in Figure~\ref{fig:heavy-tailed-distribution} and Figure~\ref{fig:light-tailed-fitting} which illustrate linearity when placed on a log-log scale. Recall that a power law model is one of the form $y = kx^n$, i.e., in which the dependent variable $y$ is proportional to some fixed power of the independent variable $x$; on a log-log scale this relationship becomes $\log_{10}(y) = n \log_{10}(x) + \log_{10}(k)$ \cite{mitzenmacher2005brief} which is the reason why linearity is observed in that scale. To extract these quantities from the sparse traffic matrix $\mathbf{A}_t$, users may rely upon the formulas provided in Figure~\ref{fig:network_quantities} which the D4M library simplifies the calculation of, especially when working with large data sets \cite{kepner2012dynamic}.

`Brightness' of destinations and sources is a well-established measure of importance when studying the behavior of network traffic \cite{devlin2021hybrid}, quantifications of which include the aforementioned destination fan-in and source fan-out statistics, respectively. A destination may be considered `bright' if it is receiving a large amount of traffic, either measured in terms of number of sources or number of packets received from sources. As such, a destination would be considered `dim' if it were receiving a relatively small amount of traffic under either measure. Using the destination fan-in measure, Figure~\ref{fig:heavy-tailed-distribution} suggests that, while rare, `bright' destinations are nonetheless present in the network, and the consistency of the power law model across time ranges and locations shows that this is the rule rather than the exception.

\subsection{Data Fitting}

\begin{figure}[h!tbp]
    \centering
    \includegraphics[width=\linewidth]{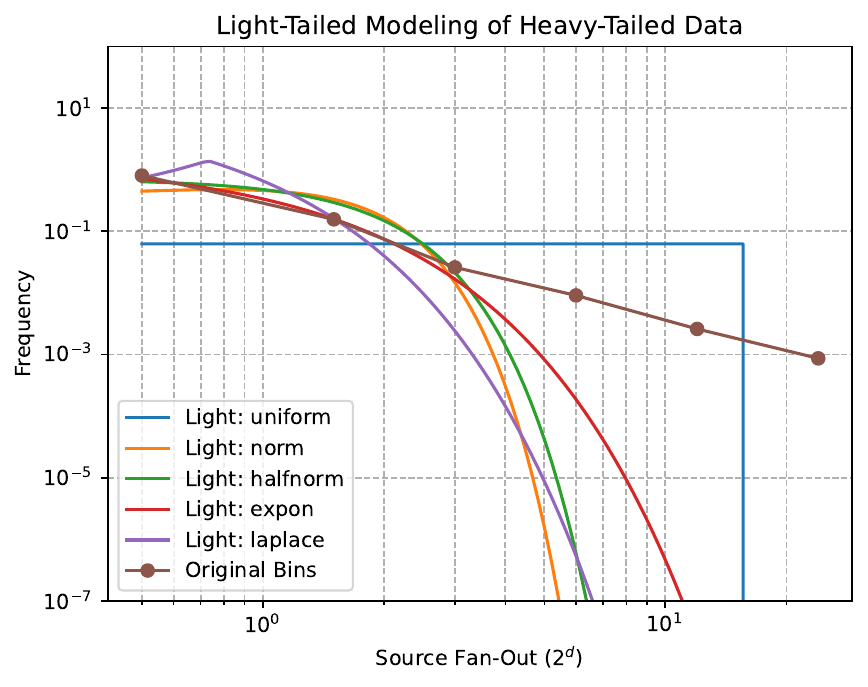}
    \caption{Model fitting of the uniform, normal, half-normal, exponential, and Laplace light-tailed distributions to the frequency distribution of logarithmically binned source fan-out quantities among all sources present in the MAWI dataset.}
    \label{fig:light-tailed-fitting}
\end{figure}

Having provided students with opportunities to see heavy-tailed distributions in action, the final section of the Lab drives home the quantitative and qualitative differences between heavy-tailed and light-tailed distributions by employing best-fit models for the destination fan-in and source fan-out distributions constructed in the previous section. An added conceptual and computational difficulty present here is the fact that---in order to best handle data spanning many orders of magnitude---logarithmic binning is needed \cite{devlin2021hybrid}. Such data is often said to be `censored' and requires specialized tools in order to produce appropriate best-fit models. SciPy's tool for handling such data is its \verb`scipy.stats.CensoredData` submodule; although relatively new, this submodule supports the needed \verb`fit` method \cite{virtanen2020scipy}. To appropriately fit the data using this function, users must first determine the amount of data within each bin, input that into the \verb`CensoredData` function, call the \verb`fit` method for different distributions to generate their respective parameters, then take the given parameters and generate the distributions on the same plot as the chosen network theoretic quantity to plot the determined `best fit' models. Figure~\ref{fig:light-tailed-fitting} shows an example of this in the case of source fan-in, specifically plotting our heavy-tailed data against the \verb`uniform`, \verb`normal`, \verb`halfnorm`, \verb`exponential`, and \verb`laplace` light-tailed distributions. As expected, despite using calculated best-fit model parameters, the tails of the light-tailed models drop off considerably in comparison to the known power-law behavior of source fan-in.

\section{Conclusion}\label{SCM}

The ubiquity of heavy-tailed phenomena across fields such as cybersecurity \cite{kepner2019hypersparse}, economics \cite{newman2005power}, social networks \cite{chakrabarti2004rmat,barabasi1999emergence,barabasi1999emergence}, and others necessitates accessible, understandable educational material. 
Across these and other domains, seemingly impossible outliers appear more commonly than the light-tailed approach would suggest, making the dependence on light-tailed tools both inappropriate and potentially dangerous. 

Analyses performed to determine the availability and usage of heavy-tailed statistics in cybersecurity and statistics education at the college-level and above showed little to no support for these ideas except in isolated cases relegated to advanced actuarial programs. Based on this, the Traffic Matrix Lab was created to provide educational material geared towards cyber operators both in terms of usefulness and in terms of accessibility, teaching users about heavy-tailed phenomena, how to identify such phenomena, and how to appropriately fit distributions to the data resulting from those phenomena. This heightened understanding can lead to improved resource allocation, anomaly detection, and risk assessment.  

Future work includes targeted roll-outs of the Traffic Matrix Lab to evaluate whether the Lab succeeds and fails in its goals, to subsequently iterate and expand on the Lab's material, and to eventually make the material publicly available so it can be leveraged by any and all cybersecurity programs, increasing the effectiveness of cyber operators and thus Internet security and privacy.

\section*{Acknowledgment}

The authors wish to acknowledge the following individuals for their contributions and support: W. Arcand, W. Bergeron, D. Bestor, C. Birardi, B. Bond, S. Buckley, C. Byun, G. Floyd, V. Gadepally, D. Gupta, M. Houle, M. Hubbell, M. Jones, A. Klien, C. Leiserson, K. Malvey, P. Michaleas, C. Milner, S. Mohindra, L. Milechin, J. Mullen, R. Patel, S. Pentland, C. Prothmann, A. Prout, A. Reuther, A. Rosa , J. Rountree, D. Rus, M. Sherman, C. Yee.

\bibliographystyle{IEEEtran}
\bibliography{references}

\end{document}